# The Cosmic Quarks

Bikash Sinha
Variable Energy Cyclotron Centre
1/AF Bidhannagar, Kolkata 700064, India
bikash@vecc.gov.in

**Abstract:**
There are at least three sources of cosmic quarks in the universe. One, the quark nuggets which may survive beyond a certain baryon number during the phase transition from quarks to hadrons microseconds after the big bang. These quark nuggets can very well be candidate of cold dark matter and these nuggets consist of strange quarks.

Second, the interior of the neutron star may well be made of quarks due to very very high pressure. It is further shown that the interior of heavy neutron star, recently discovered $\approx 2 M_\odot$, with an appropriate equation of state, can also be made of quark core.

Finally, using the property of colour entanglement among quarks it is entirely possible to have free orphan quarks roaming around in the cosmos. Some better understanding of dark energy and dark matter is possible with these entangled orphan quarks.

***Keyword:*** *cosmic quarks, quark hadron phase transition, neutron star, orphan quark*

## Introduction:

Microseconds after the Big-Bang the universe consisted of quarks, leptons and photons. As the universe attained a temperature around (150-200) MeV (as per the standard model) the quarks made a phase transition to hadrons. The nature of the phase transition however still remains terra incognita by far and large.

For very small baryon number µ → 0 and at a temperature of (T≈ 150-200 MeV) the wisdom of lattice suggests a "cross over" from the universe of quarks to the universe of hadrons. In such a situation any imprint of the epochs in the universe before the cross over is erased out!

We suggest in this paper as others have done [1,2,3] that with supercooling and consequent mini inflation the universe indeed can go through a first order phase transition leaving remnants of that epoch.

Some of the quarks nuggets which are the remnants under some circumstances can survive even up to the present time. These quark nuggets can be candidate of the cold dark matter. We discuss this point at length, as one of the candidate of cosmic quarks.

Second, is the case of neutron star, the core of which, although cold∼ 1 MeV or so, is under severe pressure (very high µ) turning the neutrons in this case, to quarks. It has been demonstrated that even for large neutron star ∼ $2M_\odot$ or so the core can be made of quarks.

Finally during the hadronisation in the microseconds old universe all the hadrons are not expected to be exact multiple of 3 (n, number of hadrons)(nx3) (explained in the last section in more detail) quark system. Not all the quarks thus do not turn to hadrons leaving individual [6] quarks alone. It will be shown that this is not anyway in contradiction of the standard model. We discuss these "orphan quarks" in this paper.

A further crucial point often overlooked is that these quarks are quantum mechanically colour entangled. The consequences of the quantum entanglement will be discussed in this paper.

## 2. Cold dark matter and the cosmic phase transition

Over the years abundant evidence has been accumulated indicating the presence of large quantities of unseen matter surrounding normal galaxies including our own [1,2,3]. The nature of this dark matter, however, remains unknown except that it cannot be normal, such as the matter in stars, gas or dust.

The evidence of the existence of dark matter is primarily gravitational. The discrepancy between the luminous mass and the gravitational mass clearly indicates the presence of a large amount of unseen mass of the universe, usually referred to as dark matter. By now, it is also established that dark matter accounts for approximately 27% of the energy budget of the universe.

All kinds of speculations are in circulation. Axions, massive neutrinos, other kind of yet unknown exotic particles, and even weakly interacting massive particles, referred to usually as WIMPs have been proposed as candidates of dark matter, but none of which are experimentally observed so far. Alcock et al. [4] and Aubourg et al. [5] (using gravitational microlensing method suggested by Paczynski [6]) discovered the existence of this dark matter. Alcock et al. [4] suggested that the dark matter can be explained by normal matter, known collectively as Massive Compact Halo Objects (MACHOs) [4]. They went on to speculate that MACHOs might be brown dwarfs, Jupiter like objects, neutron stars, or even black holes. It has been argued by us [8,9,10,11] quite exhaustively, that a natural explanation as mentioned already will be that the MACHOs are the relics from the putative cosmic phase transition from quark to hadrons about a microsecond after the Big Bang. MACHOs, it is argued, can be the quarks nuggets, which have survived from that primordial epoch. It is entirely plausible that these relic quark nuggets are made of strange matter, the true ground state of QCD, originally suggested by Witten [12].

This assertion has acquired further credibility from the recent experimental observations of the Bose Institute group [13,14] engaged in studying cosmic strangelets at a mountain altitude. Although heavy- ion collision is unlikely to produce strangelets, strange quark nuggets in the form of strangelets hurtling down through the cosmos will tend to pick up mass [13] from the atmospheric atoms, as they fall on the earth. The analysis of Banerjee et al. [13] indicates that a strangelet with an initial mass ~ 64 amu and charge ~ 2 (typically for strangelet Z/A << 1) acquires mass as it passes through the atmosphere, and goes up to, a mass ~ of 340 amu or so, at the end of its cosmic journey. This is at an altitude of ~ 3.6 km above sea level, typically the Himalayan mountain region in India such as Darjeeling. The collected data, just mentioned, suggest the interpretation of exotic cosmic ray events of very small Z/A, arising from the Strange Quark Matter droplets.

Recently, Harvey et al. [15] have suggested that collisions between galaxy clusters provide a test of the non-gravitational forces acting on dark matter. Using the Chandra and Hubble Space Telescopes, they observe 72 collisions, including both 'major' and 'minor' mergers. Combining these measurements statistically, they find the existence of dark mass of 7.6 σ significance. The position of the dark mass has remained closely aligned, implying a self-interaction cross section $\sigma_{DM}/m$ < 0.47 cm$^2$/g and disfavoring proposed extensions to the standard model.

This observation [15] is in general agreement, it seems with the identification of dark matter with strange quark nuggets, relics of the quark-hadron phase transition. It is interesting [10] to observe that going over from a radiation dominated universe to gravity dominated universe these strange quark nuggets begin to clump around the temperature ~ 1MeV of the universe. It was found by us [10] that after clumping is over, these objects feel only the influence of gravity as has been observed by Harvey et al. [15]. Thus, SQNs are in some sense the WIMPs, massive but weakly interacting, by gravity only.

Armed with the recent observations [13,14] of strangelets at mountain top and the very first observations of MACHO collaboration at Mount Stromlo by Alcock et al. [4] and by Aubourg et al. [5] of EROS collaboration at La Silla, Chile, it is proposed to go through a critical analysis of the origin of Strange Quark Matter (SQM) and it's survival through the cosmological time scale, and, finally identifying SQMs as the candidates for cold dark matter (CDM) which may indeed close the universe [10].

There are two central issues: one the formation of quark nuggets immediately after the quark-hadron phase transition and second, their survival from that primordial epoch to now.

Since the chemical potential of the universe at that point of time was close to zero and the temperature was around 200 MeV, the wisdom of lattice will lead the universe to cross over as mentioned before to the hadrons with no relic of QCD phase transition. It should be noted however that reliability of lattice calculation with bare quarks in an expanding universe is a non trivial issue.

In the next section, I analyze the circumstances precipitated essentially by some degree of supercooling accompanied by a "mini inflation". With $\mu/T \sim 1$, the entry point of the universe of quarks to the world of hadrons is shifted along the phase boundary, and a first order phase transition takes place. In the following, I analyse this scenario critically.

**Cosmic Phase Transition from Quark to Hadrons**

It is conventionally assumed that the baryon asymmetry $\eta = (n_B - n_{\bar{B}})/\gamma$ at that primordial epoch of phase transition is the same as that of today's universe $\eta \sim 10^{-10}$. There are however reasonably straightforward arguments [12,16,17] that $\eta$ at that epoch is much higher and indeed of the order of $\eta \sim \mathcal{O}(1)$ unity. However, it is seen that after the phase transition $\eta$ goes back to $10^{-10}$, as it is today. The consequence of such a possibility is discussed in the following.

Witten [12] and others [8,11,16,17,18] have argued that a first order phase transition is plausible with a "small" supercooling. In a recent private communication Witten [19] further asserted that if $n_B \approx n_{\bar{B}}/\gamma$ remains $\sim 10^{-10}$ at the point of q-h phase transition, then supercooling is implausible. However, he also points out [19] that if the baryon to photon ratio is not so small during the QCD phase transition and becomes small because of some phenomena at later times, then supercooling is plausible in principle. In the following, we demonstrate that this is entirely possible.

Another way of arguing the case is as follows.

The resolution of the issue as to whether neutrinos are predominantly Majorana fermions, as happens to be the common prejudice currently, can be decided by the currently ongoing experiments on neutrinoless double beta decay. If, contrary to the existing belief, such experiments happen to yield null results, and neutrinos are confirmed to be Dirac fermions, this scenario of baryogenesis loses its prime attraction, entailing unsavoury fine tuning.

Given such a volatile situation, alternative scenarios of baryogenesis cannot be ruled out. Prominent among these is the nonthermal Affleck-Dine mechanism [20].

The Affleck-Dine mechanism [20] has the potential to produce a baryon asymmetry of (1) without requiring superhigh temperatures. However, the observed baryon asymmetry of ($10^{-10}$) at CMB temperatures needs to emerge naturally from such a scenario. This is what is achieved through a "little inflation" of about 7 e-folding occurring at a lower temperature, which may be identified with the QCD first order phase transition [16]. Such an inflation naturally dilutes the baryon photon ratio to the observed range, even though the baryon potential before the first order phase transition may have been high (of (1) in photon units). Comparing this "little inflation" with the more standard Guth's inationary model [21], one finds that the patterns of entropy variation in the two cases are very different. In the standard inflationary model [21] the entropy is conserved during exponential expansion, and increases, due to reheating when bubbles collide, at the end of the transition. However in Guth's (standard) scenario, supercooling is very large, though, and entropy is constant. In the little inflation scenario for the case of quark-hadron phase transition, on the other hand, the entropy is constantly increasing during the quark-hadron phase transition.

The possibility and the criterion of a mini-inflationary epoch can be demonstrated in a simple way within the Friedman model of a spatially flat universe, which is homogeneous and isotropic along with an appropriate equation of state (EOS). Let the scale factor be R with an energy density $\epsilon$, and then the Friedman equation reads with $\dot{R} = \frac{dR}{dT}$,

$$\dot{R} - CR\sqrt{\epsilon} = 0 \tag{1}$$

$$\dot{\epsilon} - 3(\dot{R}/R)(\epsilon + P) = 0 \tag{2}$$

with $C = (8\pi/3)^{1/2}/M_p$ the Planck mass $M_p = 1.2 \times 10^9$ GeV. The corresponding equation of state, relating energy density $\epsilon$ and the pressure p using the bag model reads for QGP

$$\epsilon_{qg} = (37\pi^2/90)T^4 + B \tag{3}$$

$$p_{qg} = (\epsilon_{qg} - 4B)/3 \tag{4}$$

$$p = p_{qg} + p_{bg}; \epsilon = \epsilon_{qg} + \epsilon_{bg}; \epsilon_{bg} = 3p_{bg} \tag{5}$$

$$p_{bg} = 14.25\pi^2 T^4/90 \tag{6}$$

The cosmic evolution will be an inflationary one if the expansion is accelerated, $\ddot{R} \geq 0$ which leads to using equations (1) and (2)

$$\ddot{R} = -C^2 R(\epsilon + p)/2 \geq 0 \tag{7}$$

and

$$3p + \epsilon < 0, \tag{8}$$

with the solution

$$\epsilon = B \coth^2[2C\sqrt{B}(t - t_c) + arcth(\sqrt{\epsilon_c/B})], \tag{9}$$

$$R = \sinh^{1/2}[2C\sqrt{B}(t - t_c) + arcth(\sqrt{\tfrac{\epsilon_c}{B}})\sinh^{-\tfrac{1}{2}}(arcth\sqrt{\epsilon_c/B})] \tag{10}$$

at t≫$t_{exp}$ (expansion time scale for mini inflation)= $(2C\sqrt{B})^{-1}$; the expansion proceeds exponentially as $R \propto \exp(C\sqrt{B}t)$; this is mini inflation. It is worthwhile to note that in the present case, $R \propto \exp(C\sqrt{B}t)$, where $C = (8\pi/3)^{1/2}/M_p$; for Guth inflation [21] $R \propto \exp(\chi t)$ and $\chi = \sqrt{(8\pi/3)G\rho_0}$, $\rho_0$ is the initial energy density of the universe.

Guth's inflation involves the gravitational constant, whereas mini-inflation involves the Bag constant; supercooling in the standard inflation model of the universe is by 28 or more orders of magnitude, whereas here in the quark hadron phase transition it is only 7 order of magnitude.

Thus in the cosmological context, phase transition seems to be intimately connected with supercooling.

Equations (3) to (8) are satisfied for T < Ti with $T_i \simeq 0.5B^{14}$.

For temperatures below $T_0=0.65B^{1/4}$, the pressure becomes negative leading to acceleration of the universe. This is exactly what is achieved by the mini-inflation Fig. 1.

The temperature drops primarily as per the standard $t^{-1/2}$ law. Then it increases only slightly with the release of latent heat as quarks go to hadrons and the degrees of freedom are quenched. Finally T again decreases as $t^{-1/2}$ again.

The scale factor R evidently does not follow the standard $t^{1/2}$ law of the standard model.

As shown in Fig. 2, the guiding equation for mini-inflation reads as $R \propto \exp(C\sqrt{B}t)$; here the scale factor grows exponentially. In more general terms, the scale factor R gets multiplied by $\sim 10^4$, where T only decreases by only $\sim$ 70 MeV and the entropy increases by $3\times 10^9$ times (see Fig. 2). Thus in the "little inflation" scenario, the entropy is constantly increasing during the quark-hadron phase transition and the product R(t)T(t) does not remain constant where R is the scaling factor.

The increase in entropy changes $n_q/n_\gamma$ radically with dramatic consequences. The quark number decreases as $R^{-3}$, and $n_\gamma \propto T^3$, and thus the ratio $n_q/n_\gamma \propto (RT)^{-3}$ is proportional to the inverse of the entropy. More specifically, the scale factor R is multiplied by $\sim 10^4$ where T only decreases from (say) 90 to 18 MeV. The product RT is multiplied by more than $1.4\times 10^3$, and thus the entropy increases by $3\times 10^9$.

In conclusion for this part of argument, the entropy is multiplied by $10^9$ during the quark hadron phase transition, while the baryon asymmetry is divided by the same factor. Thus we get back $n_q/n_\gamma \sim 10^{-9}$ at the end of the phase transition, as required for prediction of standard primordial nucleosynthesis. It also means, for $T > T_c$, the ratio $n_q/n_\gamma \sim 1$, but we get back our first order phase transition as just mentioned. This result contrasts sharply with normal adiabatic expansion in which baryon asymmetry does not change and remains as $n_q/n_\gamma \approx 10^{-9}$.

To recapitulate, the universe is assumed to begin with a large baryon chemical potential acquired through an Affleck-Dine [20] type of mechanism. It then undergoes a period of inflation, Fig. 1, crossing the QCD first order phase transition line, while remaining in a deconfined and in a chirally symmetric phase. The universe is then trapped in a false metastable QCD vacuum state.

The delayed phase transition then releases the latent heat and produces concomitantly a large entropy density, which effectively reduces the baryon asymmetry to currently observable values. It then enters a reheating phase all the way up to the usual reheating temperature with no significant change in the baryon potential, and then the universe follows the standard path to lower temperatures.

Experimental observations of Alcock et al. [4], Aubourg et al. [5] and more recently by the Bose Institute group [13,14] are the indirect proof of such a scenario.

Finally, as suggested by Witten [19] stable quark lumps (nuggets) are extremely optimistic. It would be very lucky to fin them. The relics as pointed out in this paper are the acid test of that luck. The recent observations [13,14] along with the old MACHO observations [4,5] seem to survive the acid test. <u>Thus Cold Dark Matter is the relic of the cosmic phase transition in the form of SQM</u>.

**Survivability of Cosmological Strange Quark Nuggets (SQN)**

In Ref. [11] the present author presents a detailed discussion on the nature of the phase transition and about the survivability of SQNs. Tracing the history from Alcock and Farhi (see Ref. [11]) onwards to Madsen, Heiselberg and Riisager [22], and then to the detailed work of Bhattacharya et al. [8] using Chromo Electric Flux Tube (CEFT), we come to the conclusion that QNs with baryon number $\geq 10^{39}$ - $10^{40}$ will indeed be cosmologically stable. It is thus very relevant to ask what fraction of the dark matter could be accounted for by the surviving QNs. To put it yet another way what we wish to address in this paper is whether the cosmological dark matter, accounting for 90% or more of all the dark matter in the universe, can be made up entirely of QNs?

As per ref. [9,10,11], the universe is closed by the baryonic dark matter trapped in QNs, and we should have

$$N_B^H(t_p) = N_B^{QN} n_{QN} V^H(t_p), \qquad (11)$$

where $N_B^H(t_p)$ is the total number of baryons required to close the universe ($\Omega_B = 1$) at $t_p$, $N_B^{QN}$ is the total number of baryons contained in a single QN, and $V^H(t_p) = (4\pi/3)(ct_p)^3$ is the horizon volume. With v/c ≈ 1/√3, $n_{QN}$, the number density of Quark nuggets [11].

$$N_B^{QN} \leq 10^{-4.7} N_B^H(t_p) \tag{12}$$

As per standard Big Bang Nucleosynthesis (SBBN) $\eta \equiv n_B/n_\gamma (10^{-9}, 10^{-10})$ for convenience (as an estimate), the baryon number is well within the horizon limit at the QCD epoch ≈ $10^{49}$.

These usual baryons constitute only ~ 10% of the closure density ($\Omega_B$~ 0.1 from SBBN), a total number of $10^{50}$ within the horizon at a temperature of ~ 100 MeV would close the universe baryonically, provided these baryons do not take part in SBBN, and a criterion fulfilled by QNs. This would require $N_B^{QN} \leq 10^{45.3}$, clearly above the survivability limit of QNs.

**Neutron Star**

The mini-bang between two nuclei, although mimicing the big bang of the universe, is somewhat different. This is discussed in detail elsewhere [11].

It is clear however, that [23] the role played by the Newtonian constant of Gravity, G, in the big bang is somewhat analogous to the role played by the vacuum energy density often referred to as the bag constant, B. The Big Bang is a display of gravity, space, and time, whereas the little bang is essentially dealing with confinement and subsequently deconfinement in extreme conditions.

On the other extreme end of the phase diagram lies a domain of very high baryon density at rather low temperature, the scenario of neutron star matter, of compressed baryonic matter (CBM) at almost zero temperature. It will be of some interest to compare and contrast the "perfect fluid" property of the quark matter in the microsecond universe with the "perfect fluid" of the core of the neutron star.

It is clear however that for the early universe, we have depleting quark matter as hadronisation progresses and the universe expands in space and time. From the canonical value of $\eta/s \leq \frac{1}{4\pi}$, for quarks with hadronisation, $\eta/s$ will go on increasing as pointed at Roy et al. [24].

Eventually, the SQNs will be floating in a dilute hadronic fluid, which is not so perfect, facing more viscous drag than its quark matter counter part.

In the case of the neutron star, however, the scenario is opposite; more hadrons will be transformed to quarks, so $\eta/s$ will decrease towards the canonical value $\eta/s \leq \frac{1}{4\pi}$. For the neutron star, an approximate estimate of $\eta/s \sim T\lambda_F C_s$ will indicate [24] that with very low value of $\lambda_F \equiv (\rho\sigma)^{-1}$, a very high $\rho$ and an extremely low temperature $\eta/s$ for the quark core of the star (with $c_p \sim 1/\sqrt{3}$ (say)) may well go down below the generic value $\frac{1}{4\pi}$, close to zero, making the core, a perfect fluid splashing on the membrane [12] of hybrid hadronic matter and quark core. The recent observation of two massive neutron stars [26] M=1.97± 0.04, $M_\odot$ and M=2.01± 0.04$M_\odot$ ($M_\odot$ solar mass) indicates that equation of state of dense cold matter must be rather stiff and, that immediately raises the question quark matter, can it be stiff enough to support such massive stars?

As shown by Kojo [27], Hatsuda [28] and others that with reasonable strength of the vector repulsion between quarks and the diquark pairing interaction can readily support high mass neutron star. This scenario is consistent with other neutron stars of radii 10-12 kms. It should be noted that dense nuclear matter with three- body forces and so on do not quite describe these massive stars. How far this scenario is consistent with lattice calculations is yet to be seen.

In the Hawking- Unruh radiation scenario while 'G' ensures gravitational attraction and `B' ensures confinement [23]. Similarly, the Hawking radiation from a black hole, with its celebrated connection to entropy [25], has an analogous scenario to that of the quark nuggets that survive the cosmic phase transition. The strange quark nuggets radiate neutrons but remain dark and cold (non-relativistic). Strange Quark Nuggets are also somewhat analogous to black holes, tend to absorb matter as they hurtle through the cosmos [14]. An attempt is being made here to find out a modified entropy "entropy equivalence" between black hole and SQNs. This was first initiated by Hawking-Unruh radiation [29].

For SQNs, the deciding role is played by the Bag Pressure, B, making the SQNs dark, veiled to an exterior observer. The natural length scale for SQNs, heuristically, can be argued as $L_B = \frac{M}{M_\odot}(B^{1/4})^{-1}$, so that the entropy of SQNs is $S_{QN} = \frac{A}{4L_B^2} = k\pi R^2 B^{1/2}(M_N/M_\odot)^{-2}$, $M_N$ being the mass of the quark nugget and $M_\odot$ is the solar mass, whereas, for the black hole, $S_{BH} \equiv \frac{A}{4L_p^2} = \frac{c^3 A}{4G\hbar} : A = 16\pi(GM_B/C^2)^2$. Thus, $S_{BH} = \frac{4\pi GM_B^2}{c\hbar}$, where $M_B$ is the mass of the black hole. At this preliminary stage, the above ansatz of SQNs' entropy seem to broadly agree with other results.

As expected the ratio $S_{BH}$ to $S_{QN}$ primarily depends on $M_B$ and $M_N$,

$$\frac{S_{BH}}{S_{QN}} = \frac{4GM_B^2}{\hbar c R^2 B^{1/2}} \frac{M_N^2}{M_\odot^2}. \tag{13}$$

**Outlook**

Driven by the familiar standard model and introducing a "mini- inflation" at the cosmic quark-hadron phase transition, one can precipitate a first order phase transition from quarks to hadrons.

It has been argued in this paper that it does not introduce any non-standard cosmological scenario and indeed this mini-inflation comes in quite naturally, raising the value of $\eta = n_B/n_\gamma$ substantially before $T > T_c$ and has fallen down to the usual standard model $\eta = n_B/n_\gamma = 10^{-10}$ with increasing entropy. It is argued that the MACHOs will survive the cosmological time scale and beyond a critical baryon number window of $\sim 10^{40}$. These are the candidates for cold dark matter observed some years ago [4,5] and rather more recently [13,14].

The interestingly satisfying thought lingers that we can accommodate all this in the framework of the standard model without invoking yet unobserved exotic physics. Clearly, we should look for SQM nuggets, as candidates for cold dark matter, more vigorously.

**Colour entangled orphan quarks and dark energy from cosmic QCD phase transition**

The present day astrophysical observations indicate that the universe is composed of a large amount of dark energy (DE) responsible for an accelerated expansion of the universe, along with a sizeable amount of cold dark matter (CDM), responsible for structure formation. The explanations for the origin or the nature of both CDM and DE seem to require ideas beyond the standard model of elementary particle interactions. Here we show that CDM and DE both can arise from the standard principles of strong interaction physics and quantum colour entanglement of quark hadron phase transition.

During the past few decades, several accurate astrophysical measurements have been carried out and the large amount of data collected makes it possible to test the existing different cosmological models. Based on the knowledge gleaned so far, the present consensus [30-32] is that the standard model of cosmology, comprising the Big Bang and a flat universe is correct. The Big Bang Nucleosynthesis [33] (BBN), which forms one of the basic tenets of the standard model, shows that baryons can at most contribute $\Omega^B$ ($\equiv \frac{\rho_B}{\rho_c}$, $\rho_c$ being the present value of closure density $\sim 10^{-47}$ $Gev^4$) $\sim 0.04$, whereas structure formation studies require that the total (cold) matter density should be $\Omega_{CDM} \sim 0.23$. Matter contributing to CDM is characterized by a dust-like equation of state, pressure $p \approx 0$ and energy density $\rho > 0$ and is responsible for clustering on galactic or supergalactic scales. Dark energy (DE), on the other hand, is smooth, with no clustering features at any scale. It is required to have an equation of state $p = w\rho$ where $w < 0$ (ideally $w = -1$), so that for a positive amount of dark energy, the resulting negative pressure would facilitate an accelerated expansion of the universe, evidence for which has recently become available from the redshift studies of type IA supernovae [32, 34]. For a flat universe $\Omega \sim 1$, $\Omega_{DE} \sim 0.73$ [35] implies that $\rho_{DE}$ today is of the order of $10^{-48}$ GeV$^4$.

In this paper we assume a first order cosmic QCD phase transition as has been already shown. Other crucial ansätze in our scenario include that the universe is overall colour neutral at all times, baryogenesis is complete substantially before the QCD transition epoch and that the baryon number is an integer. Together with the assumption of quantum entanglement, the above picture is capable of explaining both dark matter, orphan quarks and to an extent, dark energy. Here we will restrict ourselves mainly to the case of orphan quarks [36].

At temperatures higher than the critical temperature $T_c$, the coloured quarks and gluons are in a thermally equilibrated state in the perturbative vacuum (the quark-gluon plasma). The total colour of the universe is neutral (i.e., the total colour wave function of the universe is a singlet). It is strongly believed from QCD analysis that this thermally equilibrated phase of quarks and gluons exhibits collective plasma-like behaviour, usually referred to as the Quark-Gluon Plasma or QGP. Then, as $T_c$ is reached and the phase transition starts, bubbles of the hadronic phase, begin to appear in the quark-gluon plasma, grow in size and form an infinite chain of connected bubbles (the percolation process). This is a complimentary scenario of the usual picture of phase transition, followed in this paper. At this stage, the ambient universe turns over to the hadronic phase. Within this hadronic phase, the remaining high temperature quark phase gets trapped in large bubbles. As is well known, this process is associated with a fluctuation in the temperature around $T_c$; the bubbles of the hadronic phase can nucleate only when the temperature falls slightly below $T_c$. The released latent heat raises the temperature again and so on. It is thus fair to assume that the temperature of the universe remains around $T_c$ at least upto percolation.

The net baryon number contained in these Trapped False Vacuum Domains (TFVD) could be many orders of magnitude larger than that in the normal hadronic phase and they could constitute the absolute ground state of strongly interacting matter as has been argued by Witten [37]. The larger TFVDs with baryon number $\sim 10^{42\text{-}44}$, which are stable and separated by large distances ($\sim 300m$), would have a bearing on the dark matter content of the universe [38] as has been argued in this paper. In all these considerations, it has been tacitly assumed that in a many-body system of quarks and gluons, colour is averaged over, leaving only a statistical degeneracy factor for thermodynamic quantities. Here we argue that such simplification may have led us to overlook a fundamentally important aspect of strong interaction physics in cosmology, that is of quantum colour entanglement, still in an emergent state.

Let us now elaborate on the situation in some detail. In the QGP, all colour charges are neutralised within the corresponding Debye length, which turns out to be $\sim \frac{1}{g_s(T)T}$, where $g_s$ is the strong coupling constant. For Debye length smaller than a typical hadronic radius, hadrons cannot exist as bound states of coloured objects. The lifetime of the QGP may be roughly estimated by the temperature when the Debye screening length becomes larger than the typical hadronic radius, and formation of hadrons as bound states of coloured objects becomes possible.

The size of the universe being many orders of magnitude larger than the Debye length, the requirement of overall colour neutrality is trivially satisfied. Another condition required for the existence of the QGP would be the occurrence of sufficient number of colour charges within the volume characterised by the Debye length; otherwise the collective behaviour responsible for the screening would not be possible. For the cosmic QGP, these conditions are satisfied till temperatures ~ 100-200 MeV, the order of magnitude for the critical temperature for QCD.

So one would argue that all the physics of the QGP is contained within a Debye length. A quick estimate would show that upto $T_c$, the Debye length is less than a fermi and the total number of colour charges (including quarks, antiquarks and gluons) within the corresponding volume is greater than 10. Note, however, our emphasis on the second ansätz above, that the baryogenesis has already taken place much before $T_c$ is reached and the value of $\frac{n_b}{n_\gamma} \sim 10^{-10}$ has already been established. Firstly, it has to be realised that this baryon number is carried in quarks at this epoch, so that the net quark number ($N_q - N_{\bar{q}}$) in the universe at all times is an exact multiple of 3. The net quark number within a Debye volume then turns out to be ~$10^{-9}$. Thus, to ensure overall colour neutrality and an integer baryon number, one must admit of long-range correlations beyond the Debye length in the QGP, the quantum entanglement property [39].

Let us now consider the process of the cosmic quark-hadron phase transition from the quantum mechanical standpoint of colour confinement. As already mentioned, the colour wave function of the entire universe prior to the phase transition must be a singlet, which means it cannot be factorized into constituent product states; the wave functions of all coloured objects are completely entangled [39] in a quantum mechanical sense. This also ensures the integer baryon number condition [36]. In such a situation, the universe is characterized by a vacuum energy corresponding to the perturbative vacuum of Quantum Chromodynamics (QCD). As the phase transition proceeds, locally colour neutral configurations (hadrons) arise, resulting in gradual decoherence of the entangled colour wave function of the entire universe. This amounts to a proportionate reduction in the perturbative vacuum energy density, which goes into providing the latent heat of the transition, or in other words, the mass and the kinetic energy of the particles in the non-perturbative (hadronic) phase (the vacuum energy of the non-perturbative phase of QCD is taken to be zero). In the quantum mechanical sense of entangled wave functions, the end of the quark-hadron transition would correspond to complete decoherence of the colour wave function of the universe; the entire vacuum energy would disappear as the perturbative vacuum would be replaced by the non-perturbative vacuum.

The earlier discussions imply that in order for the TFVDs to be stable physical objects, they must be colour neutral. This is synonymous with the requirement that they all have integer baryon numbers, i.e., at the moment of formation each TFVD has net quark numbers in exact multiples of 3. For a statistical process, this is, obviously, most unlikely and consequently, most of the TFVDs would have some residual colour at the percolation time. Then, on the way to becoming colour singlet they would each have to shed (leave behind) one or two coloured quarks. This is the inherent picture of orphan quarks.

Thus, at the end of the cosmic QCD phase transition there would be a few coloured quarks, separated by spacelike distances. Such a large separation, apparently against the dictates of QCD, is by no means unphysical in this case. The separation of coloured TFVDs occurs at the temperature $T_c$, when the effective string tension is zero, so that there does not exist any long range force. By the time the TFVDs have evolved into colour neutral configurations, releasing the few orphan coloured quarks in their immediate vicinity, the spatial separation between these quarks is already too large to allow strings to develop between them; see below. (Such a situation could not occur in the laboratory searches for quark-gluon plasma through energetic heavy ion collisions as the spatial extent of the system is ~ few fermi and the reaction takes place on strong interaction time scales.) Therefore, the orphan quarks must remain in isolation. In terms of the quantum entanglement and decoherence of the colour wave function, this would then mean that their colour wave functions must still remain entangled and *a corresponding amount of the perturbative vacuum energy would persist in the universe.*

In this sense, the orphan quarks are definitely not in asymptotic states and no violation of colour confinement is involved. If we naively assume that the entanglement among these orphan quarks imply collective behaviour like a plasma, then one can estimate the corresponding Debye screening length for this very dilute system of quarks, which would still be governed primarily by the temperature. At temperatures of ~ 100 MeV, the scale being much smaller than the mutual separation between the orphan quarks, the formation of bound states of orphan quarks would be impossible.

If we naively associate an effective radius of ~ $10^{-14}$ cm (estimated from $\sigma_{qq} = \frac{1}{9}\sigma_{pp}$; $\sigma_{pp} \sim 20mb$) with each orphan quark and defining $f_q \equiv V_{colour}/V_{total}$ ($f_{q,O}$ being solely due to orphan quarks, whereas a value of 1 for $f_q$ means complete colour entanglement) we obtain $f_{q,O} \sim N_{q,O} \times (v_{q,O}/V_{total}) \sim 10^{-42} - 10^{-44}$ (where $v_{q,O}$ is the effective volume of an orphan quark and $N_{q,O}$ is the total number of orphan quarks), so that the residual pQCD vacuum energy comes out to be in the range $10^{-46}$ to $10^{-48} GeV^4$ [36], just about the amount of DE.

To conclude, we have shown that the existence of orphan quarks and CDM can be explained entirely within the standard model of particle interactions, without invoking any exotic assumption. It is remarkable that this simple picture gives quantitatively correct amount without any fine-tuning of parameters.

*Acknowledgement:* I take this opportunity to thank Sibaji Raha who introduced the concept of cosmic entanglement in the QGP sector, Horst Stocker for critically helpful and Larry McLarren for discussion.

References:

1. Trimble V A 1987 Rev. Astr. Astrophys. 25 425

2. Fich M, Tremaine S A 1991 Rev. Astr. Astrophys. 29 409

3. Griest K 1991 Astrophys. J. 366 412

4. Alcock C et al. 1993 Nature 365 621

5. Aubourg E et al. 1993 Nature 365 623

6. Paczynski B 1986 Astrophys. J. 304 1

7. Berrett D et al. 1993 Ann. N. Y. Acad. Sci. 688 619

8. Bhattacharya P et al. 1993 Phys. Rev. D 48 4630

9. Bhattacharye A et al. 2000 Phys. Rev. D 61 083509

10. Banerjee S et al. 2003 Mon. Not. R. Astron. Soc. 340 284

11. Sinha B 2014 Int. JMP A 29 1432004

12. Witten E 1984 Phys. Rev. D 30 272

13. Banerjee S et al. 1999 Phys. Rev. Lett. 85 2000

14. Basu B et al. 2015 Astro. Par. Phys. 61 88

15. Harvey D et al. 2015 arXiv:1503.07675v2 astro-ph.co.

16. Boeckel T and Scha_ner- Bielich J 2010 Phys. Rev. Lett. 105 041301

17. Borghini N et al. 2000 J. Phys. G 26 771

18. Boeckel T and Scha_ner- Bielich J 2012 Phys. Rev. D 85 103506

19. Witten E 2014 Private Communication

20. A_eck I and Dine M 1985 Nucl. Phys. B 249 361

21. Guth A H 1981 Phys. Rev. D 23 347

22. Madsen J, Heiselberg H and Riisager K 1986 Phys. Rev. D 34 2947

23. Shuryak E, Private Communication

24. Lacey R A et al. 2007 Phys. Rev. Lett. 98 092301

25. Hawking S W 1974 Nature 24 B 30; Hawking S W 1976 Phys. Rev. D 13 191

26. P. B. Demorest et al. Nature (London) 467, 1081 (2015)

27. T. Kojo, P. D. Powell, Y. Young and G. Bayan. Phys Rev D91 (2015)045003

28. G. Baym and T. Hatsuda. Prog. Theor. Exp. Phys. (2015) 031001

29. P. Castorina, D. Kharzeev and H. Satz. Thermal Hadronization and Hawking Unruh radiation in QCD. Eur. Phys. J. C 52, 187, (2007)

30. T. Tegmark, Science 296, 1427-1433 (2002).

31. S. Sarkar, Proc. EPS Int.Conf. on High Energy Physics, Budapest, 2001 (Horvath, D., Levai, P. and Patkos, A., eds.) JHEP Proceedings Section, PrHEP-hep2001/299.

32. B. Leibundgut, Ann. Rev. Astron. Astrophys. 39, 67-98 (2001).

33. C. J. Copi, D. N. Schramn, and M. S. Turner, Science 267, 192-199 (1995).

34. R. P. Kirshner, Science 300, 1914-1918 (2003).

35. H. V. Peiris et al., astro-ph/0302225.

36. S. Banerjee et al., Phys. Lett. B 611, 27-33 (2005).

37. E. Witten, Phys. Rev. D30, 272-285 (1984).

38. S. Banerjee et al., Mon. Not. R. Astron. Soc. 340, 284-288 (2003).

39. W. K. Wootters, Phil. Trans. R. Soc. Lond. A356, 1717-1731 (1998).

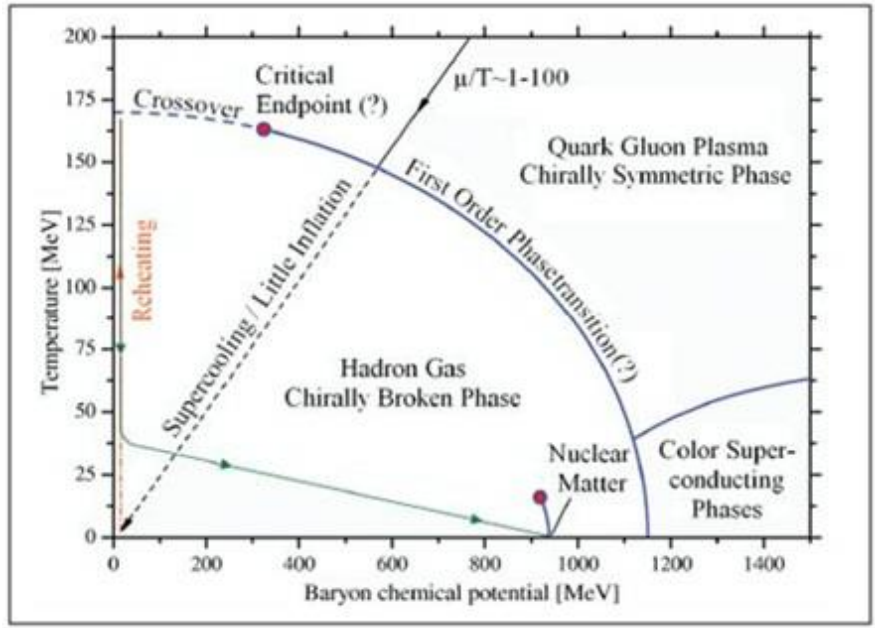

Figure 1: Sketch of a possible QCD phase diagram with the evolution path of the universe in the little inflation scenario, see Refs. [16], [18].

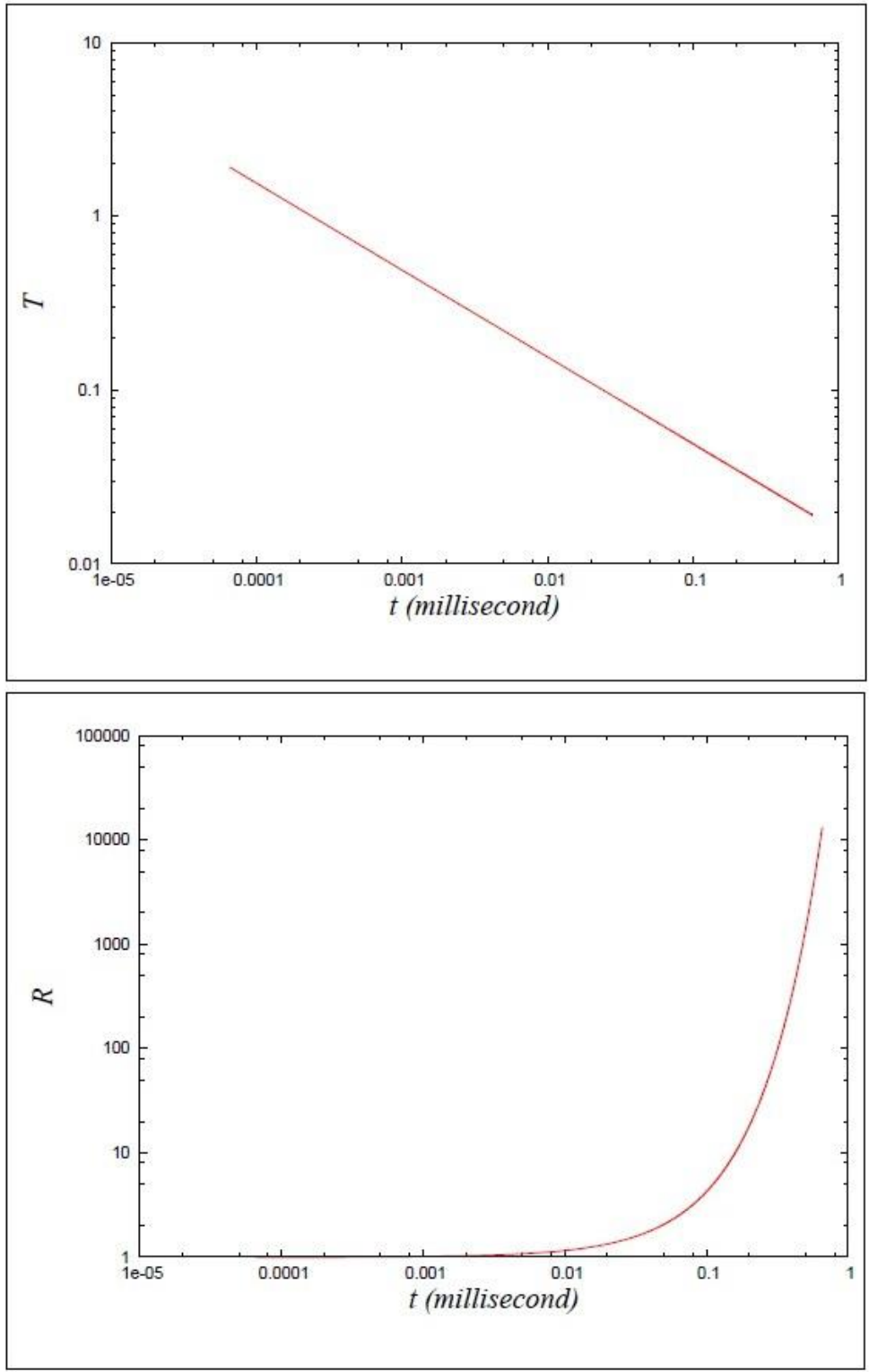

Figure 2: (Upper panel) The temperature as a function of time, Standard Cosmology. (Lower panel) The Scale factor as a function of time at the time of mini inflation: $R \propto \exp(C\sqrt{B}t), C = \sqrt{(8\pi/3)}/M_p$.